\documentclass[prc,preprint,epsfig,nofootinbib,floatfix,showpacs]{revtex4}
\usepackage{graphics}
\usepackage{epsfig}
\usepackage{amsfonts}
\usepackage{amsmath}
\usepackage{bm}

\begin{document}
\title{Giant dipole resonance in deformed nuclei: \\
dependence on Skyrme forces}

\author{V.O. Nesterenko, W. Kleinig
\footnote{Permanent address: Technische Univirsit\"at Dresden, Inst. f\"ur Analysis,
D-01062, Dresden, Germany}
}
\affiliation{Laboratory of Theoretical Physics,
Joint Institute for Nuclear Research, Dubna, Moscow region, 141980, Russia\\
nester@theor.jinr.ru; kleinig@theor.jinr.ru}

\author{J. Kvasil, P. Vesely}
\affiliation{Institute of Particle and Nuclear Physics, Charles University,
CZ-18000 Praha, Czech Republic\\
kvasil@ipnp.troja.mff.cuni.cz; vesely@ipnp.troja.mff.cuni.cz}

\author{P.-G. Reinhard}
\affiliation{
Institut f\"ur Theoretische Physik II, Universit\"at Erlangen, D-91058,
Erlangen, Germany;\\
mpt218@theorie2.physik.uni-erlangen.de}

\date{\today}

\begin{abstract}
The giant dipole resonance (GDR) in deformed nuclei is analyzed using
the self-consistent separable random-phase-approximation (SRPA) with
Skyrme forces SkT6, SkM$^*$, SLy6 and SkI3. The deformed nuclei
$^{150}$Nd and $^{238}$U are used as representative rare-earth and
actinide samples. Dependencies of the dipole strength distributions on
some basic characteristics of the Skyrme functional and nuclear matter
properties (isoscalar and isovector effective masses, time-odd
contributions) are discussed. Particular attention is paid to
the fragmentation structure of the GDR strength which are shown to
depend sensitively to spin-orbit intruder states with large angular
momentum.
\end{abstract}


\maketitle

\section{Introduction}

Skyrme forces \cite{Skyrme,Vau,Engel_75,Dob} are widely used for the
self-consistent description of ground state and excitations of atomic nuclei
(for a recent review see \cite{Ben}). Recently, there is increasing interest in
applications to the dynamics of exotic deformed nuclei, see e.g.
\cite{Stoitsov_PRC_03,Obertelli_PRC_05,Maruhn_PRC_05}. The demands on the
reliability and quality of the description are higher now then in the
pioneering earlier studies. This calls for closer inspection of the dynamical
properties of Skyrme forces. For example, there are still several open problems
related to the description of giant resonances, particularly for the for
isovector modes \cite{Rei_NPA_99}, for which the giant dipole resonance (GDR)
is the most prominent representative. Already for the description of the GDR,
there still exist a several puzzling features. For example, some trends of GDR
properties with nuclear matter characteristics look at first glance surprising
(microscopic calculations \cite{Rei_NPA_99} show a
decrease of the GDR energy with increasing symmetry energy while macroscopic
estimates predict the opposite trend \cite{Bertsch_book}). Another puzzling
feature is that some Skyrme parameterizations (like SkM$^*$) yield a too large
high-energy shoulder of the dipole strength distribution in heavy nuclei
\cite{Maruhn_PRC_05,srpa_PRC_02}.  Moreover, the influence of the effective
mass and related time-odd couplings in the Skyrme forces on the GDR spectra has
not yet been fully clarified and deserves closer inspection.
The aim of this contribution is to explore more closely these GDR properties
in heavy deformed nuclei.

Until recently, the treatment of excitations in deformed nuclei within
self-consistent models was rather involved and time consuming. Meanwhile,
a new generation of the efficient RPA schemes has emerged
\cite{Stoitsov_PRC_03,Obertelli_PRC_05,srpa_PRC_02,nes_long,prep_05},
which allow now the systematic studies.  In this contribution, we
will exploit one of these schemes, the separable random-phase-approximation (SRPA).
This method drastically simplifies the calculations and, at the same time,
provides high accuracy. It can be used for both spherical
\cite{srpa_PRC_02} and deformed \cite{nes_long,prep_05} nuclei.
Here we will apply it to study trends and spectral pattern
of the GDR in heavy deformed nuclei $^{150}$Nd and $^{238}$U.
Descriptions for four different Skyrme forces will be compared and
scrutinized.

\section{Details of Calculations}

The explicit form of the Skyrme functional used in our study is given
elsewhere \cite{srpa_PRC_02,Re92}. The calculations are performed for
the Skyrme forces SkT6 \cite{skt6}, SkM$^*$ \cite{skms}, SLy6 \cite{sly6}
and SkI3 \cite{ski3}. Though these forces were fitted with a different
bias, they all provide a good overall description of nuclear bulk
properties and are suitable for deformed nuclei (see review
\cite{Ben}). For our aim it is important that this selection of forces
covers different values of key characteristics of nuclear matter, as
shown in Table 1. We so dispose a large span of the effective
masses (isoscalar as well as isovector) and some variation of
the symmetry energy.

\begin{table}
\caption{\label{tab:skyrme}
Nuclear matter and deformation properties for the Skyrme forces
under consideration. The table represents the isoscalar effective
mass $m_0^*/m$, symmetry energy $a_{\rm sym}$, density dependence
of symmetry energy $a'_{\rm sym}=d/d\rho a_{\rm sym}$, sum rule
enhancement factor $\kappa$,  isovector effective mass
$m_1^*/m=1/(1+\kappa)$, and quadrupole moments $Q_2$ in $^{150}$Nd
and $^{238}$U.
The experimental values  of $Q_2$ are taken from \protect\cite{Goldhaber}.
}
\begin{ruledtabular}
{\begin{tabular}{@{}c|ccccc|cc@{}}
Forces & $m_0^*/m$ & $a_{\rm sym}$ [MeV]&
$a'_{\rm sym}$ [MeV\,fm$^3$] &
$\kappa$ & $m_1^*/m$ & \multicolumn{2}{c}{$Q_2$ [b]}\\
  & &  &  & & & $^{150}$Nd & $^{238}$U  \\
\hline
 SkT6  & 1.00 & 30.0 & 63 & 0.001 & 1.00 & 6.0 & 11.1 \\
 SKI$^*$  & 0.79 & 30.0 & 95 & 0.531 & 0.65 & 6.2 & 11.1  \\
 SLy6  & 0.69 & 32.0 & 100 & 0.250 & 0.80 & 5.8 & 11.0  \\
 SkI3  & 0.58 & 34.8 & 212 & 0.246 & 0.80 & 5.9 & 11.0  \\
\hline
 exp. & & & & & & 5.2 & 11.1 \\
\end{tabular}}
\end{ruledtabular}
\end{table}

The calculations employ a cylindrical coordinate-space grid with the
mesh size 0.7 fm. Pairing is treated at the BCS level. The ground
state deformation is determined by minimizing the total energy. As is
seen from Table 1, all three Skyrme forces give a reasonable
quadrupole moment in $^{238}$U. The calculated moment in $^{150}$Nd is
somewhat overestimated.  However, this nucleus is rather soft and
it is difficult to expect here a precise agreement with the
experiment. The modest overestimation is not a principle obstacle for
our study.

The dipole response involves  contributions from both time-even (nucleon
$\rho_s$, kinetic energy $\tau_s$, and spin-orbital $\Im_s$) and time-odd
(current $j_s$ and spin $\sigma_s$) densities, where $s$ denotes protons
and neutrons. Besides, the contributions
from the pairing densities $\chi_s$ is taken into account.
The contributions of the time-odd densities are driven by the variations
\begin{equation}
\frac{\delta^2 E}{\delta \vec{j}_{s_1}\delta \vec{j}_{s}}, \quad
\frac{\delta^2 E}{\delta \vec{\sigma}_{s_1}\delta \vec{j}_{s}}, \quad
\frac{\delta^2 E}{\delta \vec{j}_{s_1}\delta \vec{\sigma}_{s}}
\end{equation}
of the Skyrme functional terms \cite{srpa_PRC_02,Re92}
\begin{eqnarray}
      &&b_1 (\rho \tau - \vec{j}^2)
     - b'_1 \sum_s (\rho_s \tau_s - \vec{j}_s^2)
\\
     &-& b_4 \left( \rho (\vec{\nabla}\vec{{\Im}})
      + \vec{\sigma} \cdot (\vec{\nabla} \times \vec{j})\right)
     - b'_4 \sum_s \left( \rho_s(\vec{\nabla} \vec{\Im}_s)
              + \vec{\sigma}_s \cdot (\vec{\nabla} \times \vec{j}_s) \right)
\nonumber
\end{eqnarray}
where the total densities (like $j=j_p+j_n$) are given without the
index.  As was shown in refs. \cite{Engel_75,Dob}, the time-odd
densities naturally belong to the Skyrme functional if it involves all
the possible bilinear combinations of the nucleon and spin densities
together with their derivatives up to the second order. The time-odd
densities enter the functional only in specific combinations, as a
complement to the time-even ones, so as to keep Galilean and gauge
invariance of Skyrme forces.

The present SRPA calculations are performed in the approximation of
two generating operators, which allows to cover dynamics in both
surface and interior of the nucleus \cite{nes_long}.  The dipole
response is computed as the photo-absorption energy-weighted strength
function
\begin{equation}\label{eq:strength_function}
  S(E\lambda\mu ; \omega)
 =
  \sum_{\nu}
  \omega_{\nu} M_{\lambda\mu \nu}^2 \zeta(\omega - \omega_{\nu})
\end{equation}
where
\begin{equation}
  \zeta(\omega - \omega_{\nu})
  =
  \frac{1}{2\pi}
  \frac{\Delta}{(\omega - \omega_{\nu})^2 + (\Delta/2)^2}
\label{eq:lorfold}
\end{equation}
is the Lorentz weight with the averaging parameter $\Delta$,
$M_{\lambda\mu \nu}$ is the matrix element of $E\lambda\mu$ transition
from the ground state to the RPA state $|\nu>$, $\omega_{\nu}$ is the
RPA eigen-energy.  By using the SRPA technique
\cite{nes_long,prep_05}, we directly compute the strength function
with the Lorentz weight.  This dramatically reduces the computation time.
For example, by using a PC with CPU Pentium 4 (3.0 GHz)
we need about 25 minutes for the complete calculations of the GDR in
$^{238}$U.

The isovector dipole response is computed with the proton and neutron
effective charges $e_p^{eff}=N/A$ and $e_n^{eff}=-Z/A$, where Z, N are
numbers of protons and neutrons and A is the mass number. The
isoscalar spurious mode (center of mass) is located at 2-3 MeV and
thus is safely separated from the isovector one.  We use a large
configuration space including all proton and neutron levels from the
bottom of the potential well up to $\sim +16$ MeV.  This results in
$\sim 7300$ ($^{150}$Nd) and $\sim 9500$ ($^{238}$U) two-quasiparticle
(2QP) configurations in the energy interval 0 - 100 MeV.  In both
nuclei, the energy-weighted sum rule $EWSR=9NZ\hbar^2 e^2/(8A\pi m^*)$
is exhausted by 99$\%$, 97$\%$, 93$\%$, and 89$\%$ for SkT6, SkM$^*$,
SLy6, and SkI3 forces, respectively.

\section{Results and Discussion}

Results of the calculations are presented in Figs. 1-3.

Figure 1 exhibits the results for the dipole strength computed with a
width parameter $\Delta$= 2 MeV. This width is supposed to simulate
line broadening from nucleon escape as well as two-body collisions and
it is found to be suitable for the comparison with the experimental
data. For comparison, we show also the unperturbed dipole strength deduced
from the pure two-quasi-particle (2qp) excitations.

The figure shows
a strong dependence of the dipole response on the Skyrme
forces. This dependence is particularly pronounced for the unperturbed
strength where it is obviously related to the isoscalar effective mass
$m_0^*/m$.  Low  effective masses yield a stretched single particle
spectrum (see e.g. \cite{nest_PRC_mom}) leading to large 2qp energies.
Following this trend, the unperturbed 2qp strength in Fig. 1 exhibits
a systematic shift to higher energy from SkT6 to SkI3.

The residual
interaction in the isovector dipole channel is repulsive and moves
the dipole strength to higher energies. The corresponding collective
energy shift is determined by the isovector parameters listed in
Table 1. What is remarkable, there is a clear correlation in the
dependencies of the 2qp strength and the collective shift on the
Skyrme force. The dependencies are opposite. A  small effective mass
$m_0^*/m$ stretches the 2qp spectrum and, at the same time, we have
reduction of the residual interaction driven by $m_1^*/m$. Thereby,
both effective masses contribute and maybe even correlate. The latter
might be justified by that both effective masses originate from one and
the same term of the Skyrme functional $\sim b_1, b'_1$. After all, we
obtain similar GDR energies for all the forces, in fair agreement with
the experimental data \cite{nd_e1_exp,u_e1_exp}. These results
corroborate the experience from spherical nuclei that SRPA with Skyrme
forces provides a reasonable description of the GDR in heavy nuclei
\cite{srpa_PRC_02}.
\begin{figure}[th]
\centerline{\psfig{file=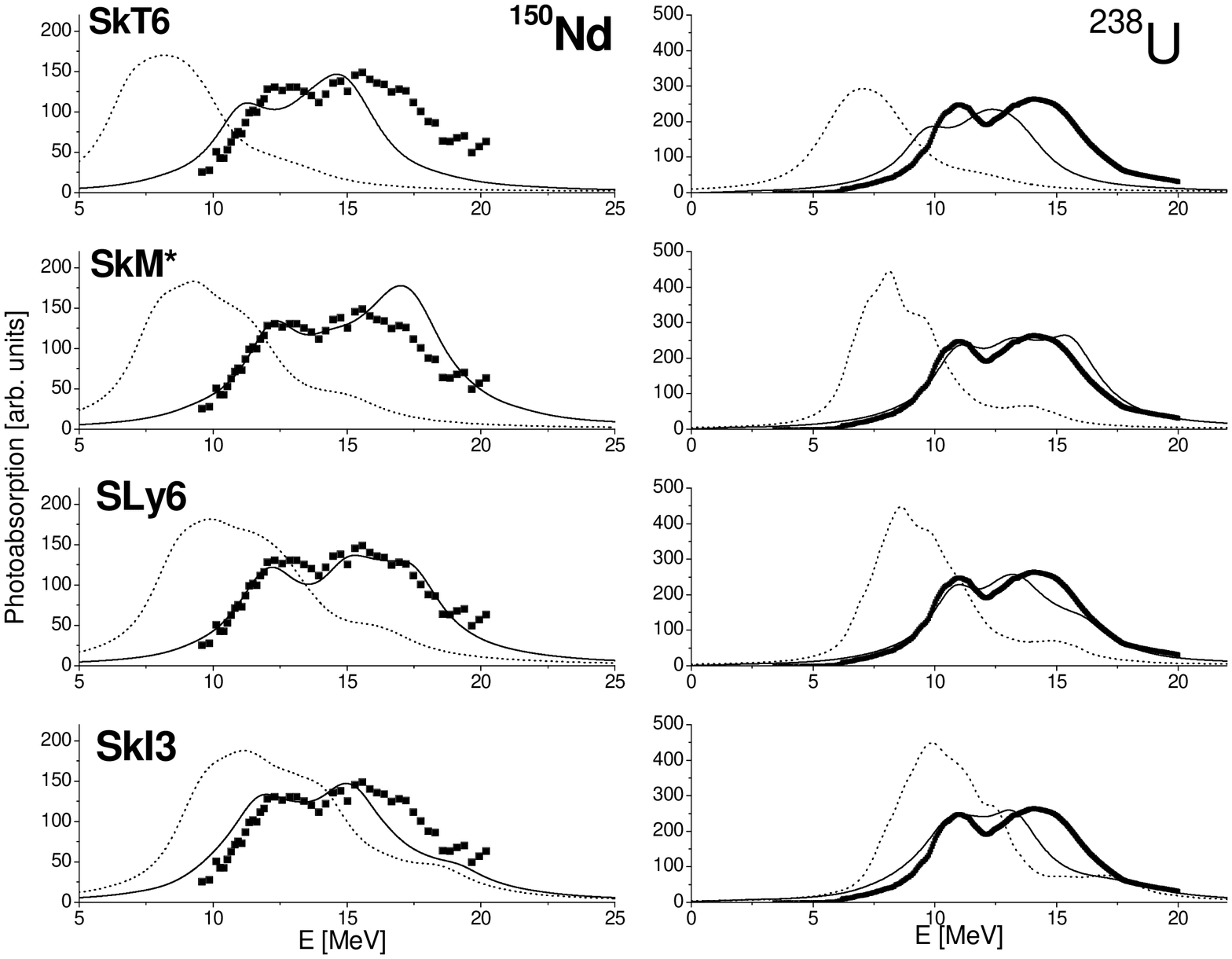,width=10cm,angle=-90}} \vspace*{8pt}
\caption{ The dipole giant resonance in $^{150}$Nd and $^{238}$U, calculated
with the Skyrme forces SkT6, SkM$^*$, SLy6 and SkI3. The calculated strength
(solid curve) is compared with the experimental data
\protect\cite{nd_e1_exp,u_e1_exp} (triangles). The quasiparticle
(unperturbed) strength is denoted by the dotted curve. The Lorentz averaging
parameter is $\Delta$=2 MeV.}
\end{figure}

Having a closer look at the full strengths in Fig. 1, we see
still some differences between SkT6, SkM$^*$, SLy6 and SkI3 cases.  There is
a small shift of the average peak position as well as different strength
patterns. The most prominent peculiarities can be explained by the
isovector parameters listed in Table 1. For example,
the exceptionally large collective shift in SkM$^*$ can be related
to very low isovector effective mass $m^*_1/m=$0.65 for this force
or, which is the same, to a high value of the sum rule enhancement
factor $\kappa =$0.531. It is seen that this case drastically deviates
from the SkT6 one where the impact of the isovector parameters
is negligible ($m^*_1/m=$1.0 and $\kappa =$0.001).

The trend of the average peak position devotes a special analysis. It
can be related to the isovector parameters in the combination
$\sqrt{a_{\rm sym,act}/m^*_1}$ where
$a_{\rm sym,act}\approx a_{\rm sym}-a'_{\rm sym}\rho_{\rm nm}/2$ is an
estimate for the actual symmetry energy in a heavy nucleus and the
density $\rho_{\rm nm}/2$ at the nuclear surface determines the GDR
response. This predicts the sequence of the peak heights
with the highest SkM$*$, smaller SLy6 and lowest SkT6
and SkI3, in agreement with Fig. 1.
Note that this explanation takes care of the ``actual''
symmetry energy whose trend deviates from the trend of the volume
symmetry energy $a_{\rm sym}$ (which then delivers the wrong trend in
macroscopic estimates of the GDR peak \cite{Bertsch_book}).

The detailed pattern of the strength distributions is caused by the
fragmentation of the bulk dipole peak over energetically close 2qp states.
Fig. 1 shows that the spectral details considerably vary with the force.
This is best visible for SkM$^*$ case where unrealistically high
right GDR shoulder and the subsequent overestimation of the resonance
width take place, especially in $^{150}$Nd. The effect
becomes weaker for SLy6 (thus giving the best description of GDR) and
vanishes for SkI3 (already with underestimation of the resonance
width).  The appearance of the right shoulder for some Skyrme forces,
preferably those with a large effective mass, was also noted in the
calculations for deformed rare-earth and actinide nuclei within the
full (non-separable) Skyrme RPA \cite{Maruhn_PRC_05} and in the SRPA
calculations for $^{208}$Pb \cite{srpa_PRC_02}. This effect seems to
be universal for GDR in heavy nuclei, independently on their shape.

\begin{figure}[th]
\centerline{\psfig{file=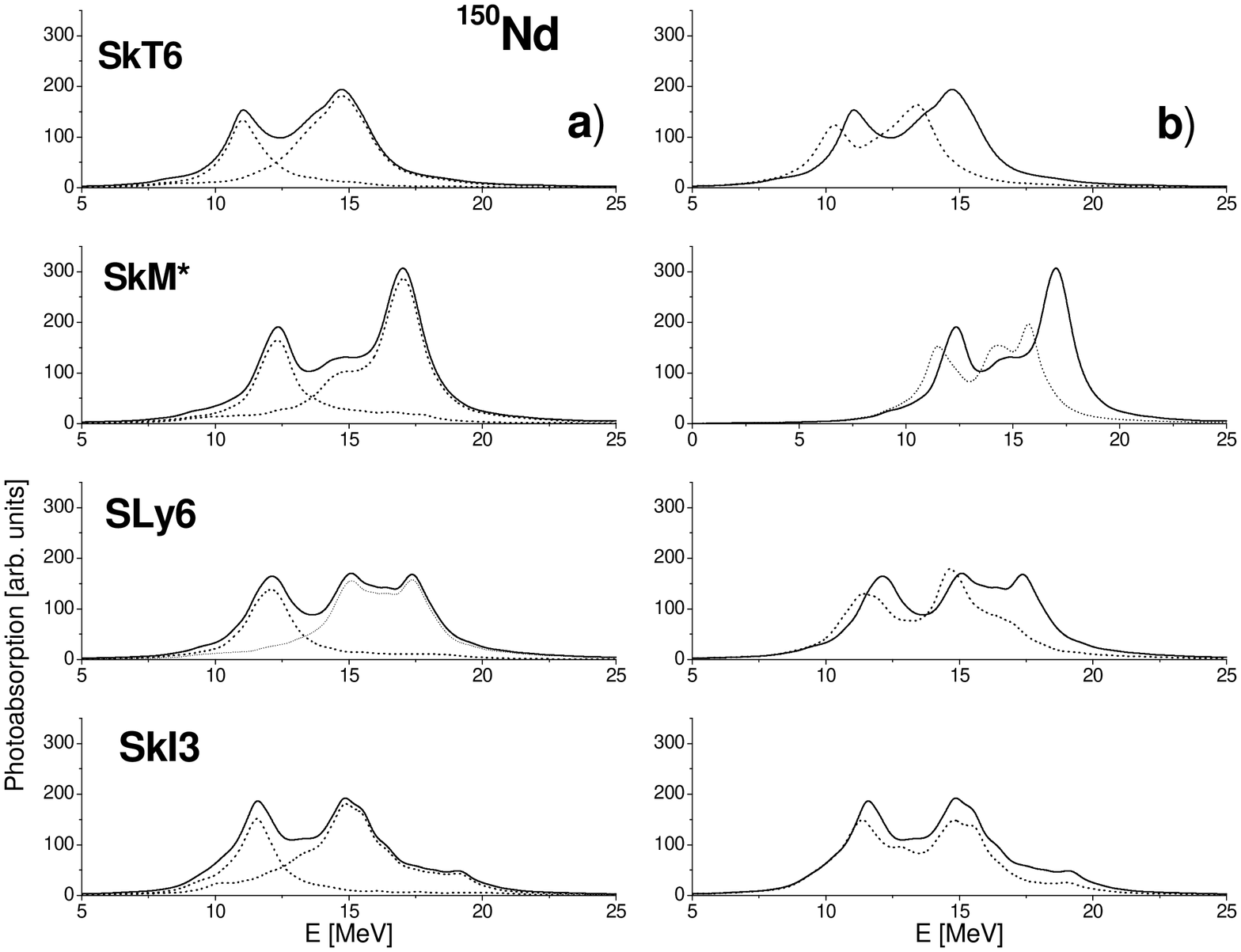,width=10cm,angle=-90}}
\vspace*{8pt}
\caption{
The dipole giant resonance in $^{150}$Nd and $^{238}$U, calculated with
the Skyrme forces SkT6, SkM$^*$, SLy6 and SkI3. Plots a) exhibit
the full strength (solid curve) and its branches $\mu = 0$
(left dotted structure) and $\mu = 1$ (right dotted structure). Plots b) exhibit
the full strength with (solid curve) and without (dotted curve) the contribution
from the proton sub-shell $1g_{9/2}$ and the neutron sub-shell $1h_{11/2}$.
The Lorentz averaging
parameter is $\Delta$=1 MeV.
}
\end{figure}

The right shoulder of the GDR is studied in more detail in Fig. 2. To
provide a more detailed description, we use here a smaller averaging
of $\Delta$= 1 MeV. The left panels of the figure show the two
GDR branches with projections $\mu=0$ and $\mu=1$.
It is seen that the right shoulder effect is not
caused by the deformation splitting since it appears only
in the branch with  $\mu =1$. Instead, it is rather a consequence of
the strength fragmentation. Following Fig. 2a, the branch $\mu =1$
undergoes a dramatic transformation
from SkM$^*$ to SkI3: its strength flows from the right to the left
flank. To understand such behavior, we should take into account that
the {\it unperturbed} dipole strength also has some kind of a right
shoulder or a strong right tail (see Fig. 1) and this may cause a
considerable fragmentation. The SkM$^*$ produces a maximal
collective shift and thus places most of the $\mu =1$ strength beyond
the tail. This minimizes the fragmentation and collects the
strength into the narrow peak.  Vise versa, the SkI3 collective shift
is small and insufficient to push the strength beyond the tail.  So in
this case the strength is fragmented between nearby 2qp
pairs and we do not observe anymore large concentration of
strength at the upper end of the spectrum.

\begin{figure}
\centerline{\psfig{file=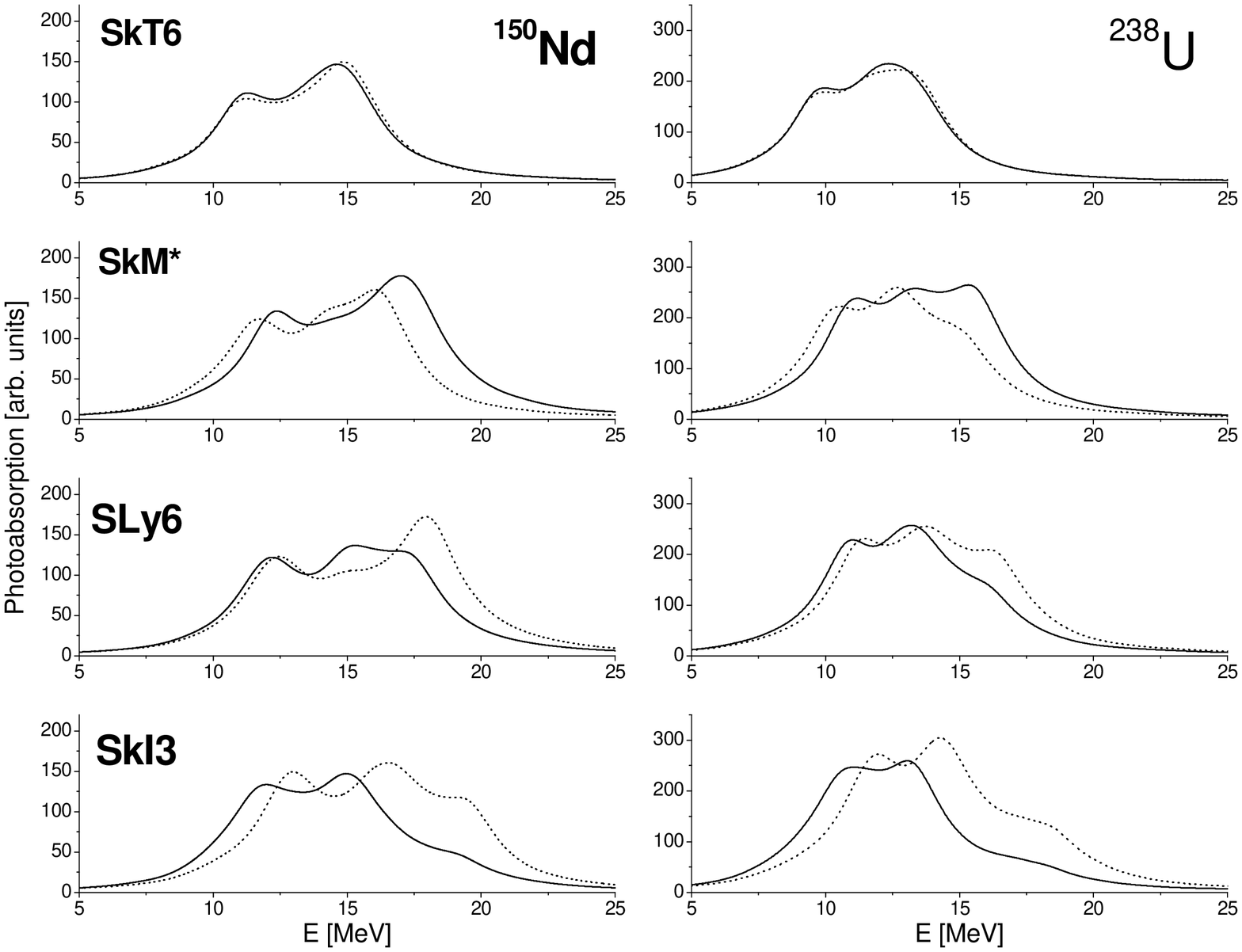,width=10cm,angle=-90}}
\vspace*{8pt}
\caption{
The dipole giant resonance in  $^{150}$Nd and $^{238}$U, calculated with
the Skyrme forces SkT6, SkM$^*$, SLy6 and SkI3. The full strength is
calculated with (solid curve) and without (dashed curve) the current
contribution. The Lorentz averaging parameter is $\Delta$=2 MeV.
}
\end{figure}

It is interesting to figure out which single-particle states are
responsible for the right flank of GDR. A simple analysis shows that
these are the occupied intruder states $j=l+1/2$ with the largest orbital
momentum $l$.
These sub-shells dive into the valence shell due to the strong
spin-orbit splitting and, the heavier the nucleus, the stronger their
impact.  In $^{150}$Nd, the intruder sub-shells are $1g_{9/2}$ for
protons and $1h_{11/2}$ for neutrons. In $^{238}$U, they are
$1h_{11/2}$ and $1i_{13/2}$, respectively. Fig. 2b shows that
excluding these sub-shells indeed weakens the right flank of GDR. It
worth noting that such sub-shells should manifest themselves in GDR of
heavy nuclei independently of the nuclear shape. The same effect
appears in spherical heavy nuclei as well.

As a next step, we consider the influence of time-odd densities in the
residual interaction.  Following our calculations, only the
current-current contribution $\delta^2 E/\delta \vec{j}_{s_1}\delta
\vec{j}_{s}$ is essential, while contributions related with the spin
density are negligible.  So, we will discuss only the effect of the
current-current term $\sim (b_1+b'_1\delta_{s,s'})$. Fig. 3 shows that
the time-odd contribution strongly depends on the force and, moreover,
obviously correlates with the differences in strength distributions
presented in Fig. 1.  This is not surprising since the squared current
density is involved to the terms $\sim b_1, b'_1$ of the Skyrme
functional and just the time-even partners in this term are
responsible for the effective masses, see Eq. (2).  As is seen from
the figure, the time-odd effect is negligible for SkT6 (where
$m^*_0/m=m^*_1/m=1$ and thus the influence of the effective masses is
minimal) but quite strong for other forces. Besides, the effect may
have different sign (compare SkM$^*$ versus SLy6 and SkI3). The
surprisingly large influence of the time-odd terms (and of the effective
mass) can be understood by the fact that the dominant contributions to
the dipole response from the principle Skyrme terms have different signs
and thus considerably compensate each other \cite{nes_long}. Hence
small effects, like the time-odd terms, acquire more weight.

It is remarkable, that the time-odd densities and effective masses
affect the same part of GDR, namely its right flank. As was shown above,
this part of the resonance strictly depends on the intruder states with a high
orbital moment. Such states should be very sensitive to any velocity
dependent values and hence to the current and effective masses. Then
it is clear why just the right GDR flank is mainly affected by these
values. This property of the GDR can be effectively used for
additional testing the isovector parameters.

Altogether, we see that the average peak position is basically
determined by the isovector parameters (symmetry energy, isoscalar
effective mass) while the detailed fragmentation pattern is also
strongly influenced by the isoscalar effective mass. A systematic
analysis of dipole strength distribution over many spherical and
deformed nuclei will help us to learn more about the underlying
single particle spectra and dynamics.

\section{Conclusions}

The giant dipole resonance (GDR) in deformed nuclei has been
investigated within the separable RPA (SRPA) method.  Four different
Skyrme forces (SkT6, SkM$^*$, SLy6, and SkI3) with different nuclear
matter characteristics (symmetry energy, isoscalar and isovector
effective mass) were applied. As the test cases, the typical axially
deformed nuclei, $^{150}$Nd and $^{238}$U, were considered.
SRPA was found to provide a good description of the GDR and, what is
important, with a minimal computational effort. This method is
indeed an efficient theoretical tool for systematic exploration of
the dynamics of deformed nuclei.

All four Skyrme forces in our sample reproduce in general the
average position of the GDR strength and its two-bump structure.
There are some trends in the peak positions (shifts about $\pm$ 1 MeV)
which can be explained by the different isovector properties of the
forces.
At closer inspection, we see considerable differences in the GDR width
and fragmentation pattern.  For example, some forces as, e.g., SkM$^*$
result in a too high right shoulder of the GDR strength distribution.
We show that height and position of this shoulder are determined by
high-angular-momentum states which are shifted down into the valence
shell by the strong spin-orbit force.  Thus many factors
(effective mass, spin-orbit force) have a significant influence on the
fragmentation pattern of the spectra, especially at the right flank of
the GDR.  This feature, in turn, provides useful information for
determination of the Skyrme parameters and
exploration of the possible correlations.

We have also discussed the effect of the time-odd terms on the GDR
profile. The time-odd spin-orbit terms have only negligible
influence. The current-current coupling, however, contributes
substantially, and the lower the isoscalar effective mass, the
stronger the contribution. This effect is crucial to
counterweight the spectral stretching of the unperturbed
excitations.

As we have seen, the detailed pattern of the resonance spectra carry worthwhile
information on the underlying single-particle spectrum of the self-consistent
mean field. The newly developed SRPA code for deformed nuclei provides access
to a much larger pool of data on nuclear giant resonances. This allows more
systematic investigations to disentangle various influences and improve
description of nuclear excitation properties. Work in that direction is in
progress.

\section*{Acknowledgements}

The work was supported  by the DFG grant  GZ:436 RUS 17/104/05,
Heisenberg-Landau (Germany-BLTP JINR) grants for 2005 and  2006 years,
and the BMBF, contracts 06 DD 119 and 06 ER 808, and by the research plan MSM
0021620834 of Clench Republic.

\end{document}